\documentstyle[epsf,aps,prl,multicol]{revtex}
\draft

\newcommand{\br}{{\bf r}}

\begin{document}
\title{Interacting Arrays of Steps and Lines in Random Media }
\author{Jan Kierfeld$^{(1,2)}$ and Terence Hwa$^{(1)}$}
\address{$^{(1)}$ Department of Physics, University of California 
at San Diego, La Jolla, CA 92093-0319\\
$^{(2)}$ Institut f\"ur Theoretische Physik der
Universit\"at zu K\"oln, D-50937 K\"oln, Germany}
\date{\today}
\maketitle

\begin{abstract}
The phase diagram of two interacting planar arrays of directed lines in 
random media is obtained by a renormalization group analysis.
The results are discussed in the contexts of the roughening of 
reconstructed crystal surfaces, and the pinning of 
flux line arrays in layered superconductors.
Among the findings are a glassy flat phase with
disordered domain structures, a novel second-order phase transition with
continuously varying critical exponents,  and the generic disappearance 
of the glassy ``super-rough'' phases found previously for a single array.

\end{abstract}


\begin{multicols}{2}

The statistical mechanics of planar arrays of directed lines is of interest
to various physical systems. For example, the steps formed on a vicinal
crystal surface can be modeled by an array of directed lines confined in a
plane~\cite{sr}. The same model describes Josephson vortex lines in a planar 
Josephson junction subject to a parallel magnetic field ~\cite{2dvg,NLS}.
An important issue concerns the behavior of such arrays of lines in the
presence of quenched disorders. It is known
that the lines are pinned by point disorders and become ``glassy'' at low
temperature~\cite{co}. However, different analytic~\cite{co,gh,vf,k,gld} and
numerical~\cite{numerics} studies have yielded conflicting results regarding
the nature of the glass phase. A subject of debate is the possible breaking
of replica-symmetry in the glass-phase~\cite{gld2,jkf}.

\begin{figure}
\epsfxsize=3.2truein
\hskip 0.0truein \epsffile{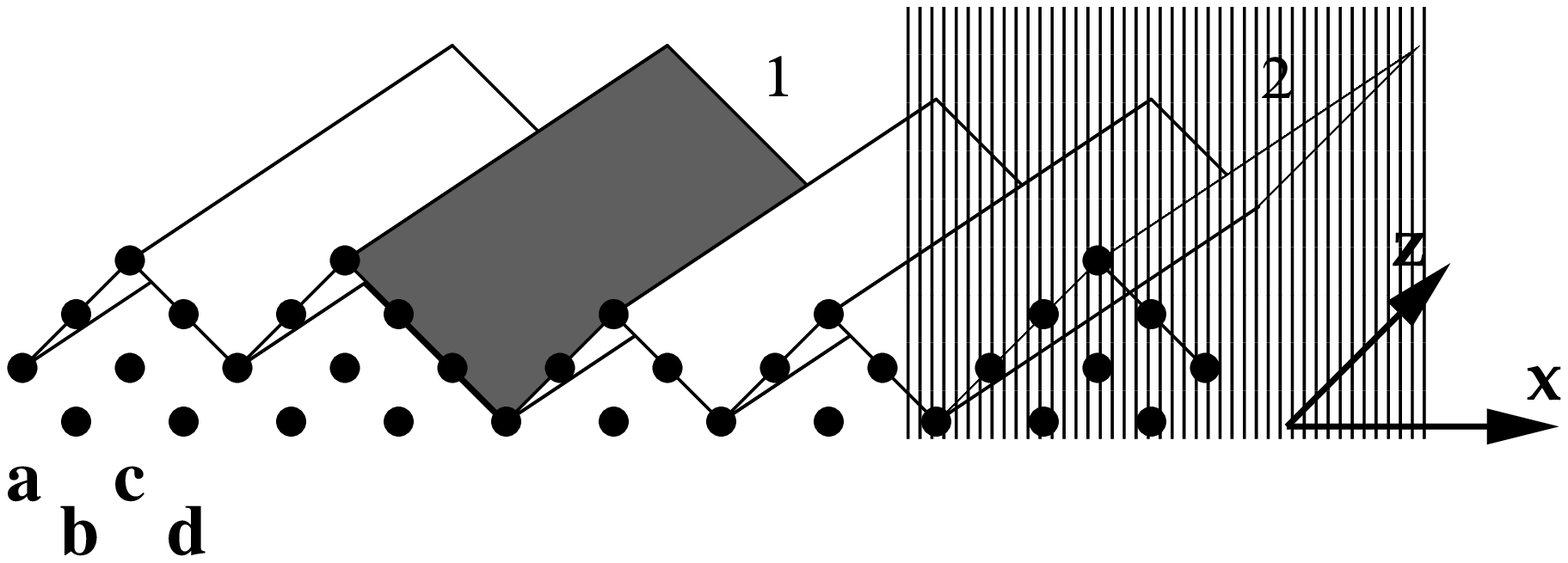}

\vspace{\baselineskip} 
 { FIG 1: Two kinds of (3x1) microfacets on a
(2x1) reconstructed crystal surface. The background (2x1) facets can be 
on four possible sublattices (marked ``{\bf a}''--``{\bf d}'').
Each (3x1) facet shifts the phase by one sublattice.} 
\end{figure}

In this article, we study the effect of point disorders on {\em two}
interacting species of lines in planar arrays. This problem arises in the
study of anisotropically (2x1) reconstructed gold (110) surfaces~\cite{vv},
where two kinds of (3x1) microfacets can be treated as two species of
interacting lines~[Fig.~1]. Previous studies of the pure system have
revealed a rich phase diagram with a variety of possible phases as a
function of the interaction parameters~\cite{vv,bk}. The inclusion of point 
disorders, say crystalline defects originating from a disordered underlying 
substrate, induces deformations in the trajectories of the microfacets.
Similar issues arise in two layers of magnetically interacting Josephson
vortex lines.
Performing a renormalization group (RG) analysis in replica space, we are
able to obtain a complete picture of the RG-flow. The result is applied to
discuss the rich phase diagrams of the anisotropically (2x1) 
reconstructed crystal surfaces. 
We also discuss the structure of the glass phases obtained
for  two planar vortex arrays, and comment on the relevance of our results
to the issue of replica-symmetry breaking in a single vortex array.

A single species of directed lines confined in a plane containing quenched
randomness can be described by the continuum Hamiltonian~\cite{2dvg,NLS} 
\begin{equation}
\beta {\cal H}_{2D}[\phi ,V]=\int_{{\bf r}}\biggl\{ \frac K2(%
\mbox{\boldmath$\nabla$}\phi )^2-V({\bf r})\rho (\phi ({\bf r}),{\bf r}%
)\biggr\}   \label{H1lay}
\end{equation}
on length scales exceeding the line spacing $l$. The first part of 
(\ref{H1lay}) gives the elastic energy of the line array in terms of a
displacement-like scalar field $\phi ({\bf r})$ 
(a displacement by $l$ corresponds
to a shift of $2\pi $ in $\phi $), characterized by an (isotropized) elastic
constant $K$~\cite{iso}. The second term describes density variations $\rho
(\phi ,{\bf r})$ induced by a random potential $V({\bf r})$. The density
field has the form $\rho (\phi ({\bf r}),{\bf r})\approx \rho _0(1-\partial
_x\phi /\left( 2\pi \rho _0\right) +2\cos \left[ 2\pi \rho _0x-\phi \right] )
$, where $\rho _0=1/l$ is the average line density, and ${\bf r}=(x,z)$ with 
$z$ along the line direction. The random potential is taken to have zero
mean with short-range correlations 
$\overline{V({\bf r})V({\bf r}^{\prime })}
=g\delta ^2({\bf r}-{\bf r}^{\prime })$ of (bare) strength $g$.

An interaction between two such species of lines
in the form of 
$ \int_{{\bf r}_1,{\bf r}_2}
V_{{\rm int}}({{\bf r}_1}-{{\bf r}_2})\rho (\phi _1,{{\bf r}_1})
\rho (\phi _2,{{\bf r}_2})$ with a short-ranged potential 
$V_{\rm int}$~\cite{int} leads to
\begin{equation}
\beta {\cal H}_{{\rm int}} \approx \int_{{\bf r}}
\biggl\{ 2\mu\rho_0^2 \cos {(\phi _1-\phi _2)}+K_\mu
\mbox{\boldmath$\nabla$}\phi _1\cdot \mbox{\boldmath$\nabla$}\phi _2\biggr\}.  
\label{Hint}
\end{equation}
with $\mu = \int_{{\bf r}}V_{{\rm int}}(\br)$
and $K_\mu =\mu /8\pi ^2$~\cite{iso}. We assume the disorder potential
$V_i$ acting on species $i$ to be statistically identical, with the
cross-correlations $\overline{V_1({\bf r})V_2({\bf r}^{\prime })}=g_\mu
\delta ^2({\bf r}-{\bf r}^{\prime })$ to be specified shortly. The full
Hamiltonian of our system, ${\cal H}[\phi _1,\phi _2]=\sum_{i=1}^2{\cal H}%
_{2D}[\phi _i,V_i]+{\cal H}_{{\rm int}}[\phi _1,\phi _2]$, can then be
written in a succinct form (after neglecting irrelevant terms~\cite{iso}), 
\begin{eqnarray}
\beta {\cal H}[\phi _1,\phi _2]= &&
     \int_{{\bf r}}\biggl\{ \sum_{i,j} \frac{K_{ij}}2%
\mbox{\boldmath$\nabla$}\phi _i\cdot \mbox{\boldmath$\nabla$}\phi _j
  -\sum_{i}{\bf w}_i\cdot \mbox{\boldmath$\nabla$}\phi _i   
   \nonumber \\
&& -\sum_{i}W_i(\phi _i,{\bf r})+
  2\mu\rho_0^2 \cos \left( \phi _1-\phi _2\right)
   \biggr\} ,  \label{H}
\end{eqnarray}
with effective random potentials $W_i$ and ${\bf w}_i$ whose correlators are 
$\overline{W_i(\phi ,{\bf r})W_j(\phi ^{\prime },{\bf r}^{\prime })}
=2g_{ij}\rho_0^2\cos(\phi -\phi ^{\prime })
   ~\delta ({\bf r}-{\bf r}^{\prime })$,
 and $\overline{{\bf w}_i({\bf r}){\bf w}_j({\bf r}^{\prime })}=\Delta
_{ij}\delta ({\bf r}-{\bf r}^{\prime })$, with the latter having the bare
values $\Delta _{ij}=g_{ij}/(8\pi ^2)$. The parameters of the theory are $%
K_{ij}=K,g_{ij}=g,\Delta _{ij}=\Delta $ for $i=j$ and $K_{ij}=K_\mu
,g_{ij}=g_\mu ,\Delta _{ij}=\Delta _\mu $ for $i\ne j$. The appearance of
the cosine couplings reflects the discrete translational symmetry of the
line arrays and has been extensively discussed in the literature.

A physical observable of interest for the reconstructed crystal
surface is the interface height profile $h({\bf r})$. It is given
by $h({\bf r})=\left[ \phi _1({\bf r})-\phi _2({\bf r})\right] /2\pi $ since
lines from species 1 and 2 represent upward and downward steps respectively,
and it is the {\em difference} of the two that determines the height profile
(see also Ref.~\cite{bk}). Note also that the (2x1) reconstructed crystal
has a choice of four possible sublattices on the surface (see Fig.~1).
Each step is thus also a domain wall separating domains of the sublattices.
An order parameter capturing the ordering of the domains is given by 
$R(\br)=\exp[i(\phi_1(\br)+\phi_2(\br))/4]$.
 To find the large scale behaviors of these observables, 
we apply the Replica method to average over disorders. 
A replica-symmetric RG analysis~\cite{dipl} for the replicated system yields 
the following 
recursion relations (to bilinear order) upon a change of scale by a factor 
$e^l$: 
\begin{eqnarray}
dg/dl &=&({\kappa }-\tau )g-g^2-g_\mu \mu ~,  \nonumber \\
dg_\mu /dl &=&({\kappa }-\tau -{\delta })g_\mu -gg_\mu -g\mu ~,  \nonumber \\
d\mu /dl &=&(2{\kappa }-{\delta })\mu -(\tau +{\kappa })g_\mu -g\mu ~, 
\nonumber \\
d{\kappa }/dl &=&\mu (\mu -2g_\mu )/2,\qquad d{\delta }/dl=(g^2-g_\mu ^2)/2~,
\label{RG}
\end{eqnarray}
where ${\kappa }=1-[4\pi (K-K_\mu )]^{-1}$ is a reduced elasticity parameter, 
${\delta }=8\pi (\Delta -\Delta _\mu )$, 
and a non-universal numerical factor 
is absorbed into $g$, $\mu $ and $g_\mu $. The RG flow is
controlled by the reduced temperature, $\tau =[4\pi (K+K_\mu )]^{-1}-1$,
which is not renormalized due to a statistical tilt symmetry.
[There exists a sixth RG equation $d\overline{\delta}/dl = (g^2 + g_\mu^2)/2$, 
for the parameter $\overline{\delta}= 8 \pi (\Delta + \Delta_\mu)$.
It however does not feedback into (\ref{RG}).]

Before delving into the structure of the RG flow, we first mention two
limiting sub-problems which have been studied previously. In the limit 
$\mu,g_\mu =0$, the elasticity parameter ${\kappa }$ is also not renormalized,
and the RG equation has the same structure as that obtained for a single
species of lines~\cite{co}, with an effective temperature $\tau_{\rm eff}
= \tau - \kappa$.
As shown in Fig.~2 (inlet (a)),
The disorder ($g$) is irrelevant at high temperatures ($\tau >\kappa $),
yielding the usual logarithmic roughness for 2D surfaces,
accompanied by a quasi-long-ranged domain order, with power law
decays in $\overline{\langle R(\br)R({\br}') \rangle}$. We refer to this
as the decoupled line (DL) phase. 
At $\tau =\kappa $, the
marginal irrelevance of $g$ yields a marginally-coupled line  phase (ML), which
again has logarithmic roughness and quasi-long-ranged domain ordering. 
At low temperatures
(${\tau <\kappa }$), the disorder is relevant. The resulting glass phases
are described by the line of fixed points $g^{*}={\kappa }-\tau $
which are perturbatively accessible for $\left| \kappa \right| ,\left| \tau
\right| \ll 1$. There, one finds the interface to be {\em super rough}~\cite
{sr}, with $\overline{\left\langle h^2\right\rangle }\sim \log ^2{L}$ (due
to the divergence of $\delta $) on large scales $L$. The glassiness is also
reflected by a disordered (short-ranged) domain order, i.e., a faster than
power-law decay in $\overline{\langle R(\br)R({\br}') \rangle}$, 
due to the divergence of $\overline{\delta}$. 
Complete decoupling for the interaction (\ref{Hint}) implies 
that $K_\mu =0$ also, or  ${\kappa }+\tau=0 $. 
Thus $g^{*}=2\left| \tau \right| $
 (and $\mu ^{*}=0,$ $\kappa ^{*}=\left| \tau\right| $) is the physical fixed
point describing a  decoupled glass (DG) phase. 
Another well known limit of our problem is that of vanishing
disorder ($g,g_\mu =0$), where a Kosterlitz-Thouless (KT) transition occurs
independent of $\tau $ (see Fig.~2 inlet (b)). 
For large coupling $\mu $, the two species become locked together, 
forming an elastically-coupled line (EL) phase with 
$\mu^{*},\kappa ^{*}\to O(1)$ (and $g^{*}=0$).
Since the up and down steps are now paired, the interface is {\em flat},
with again quasi-long-ranged domain order.

\begin{figure}
\epsfxsize=3.2truein
\hskip 0.0truein \epsffile{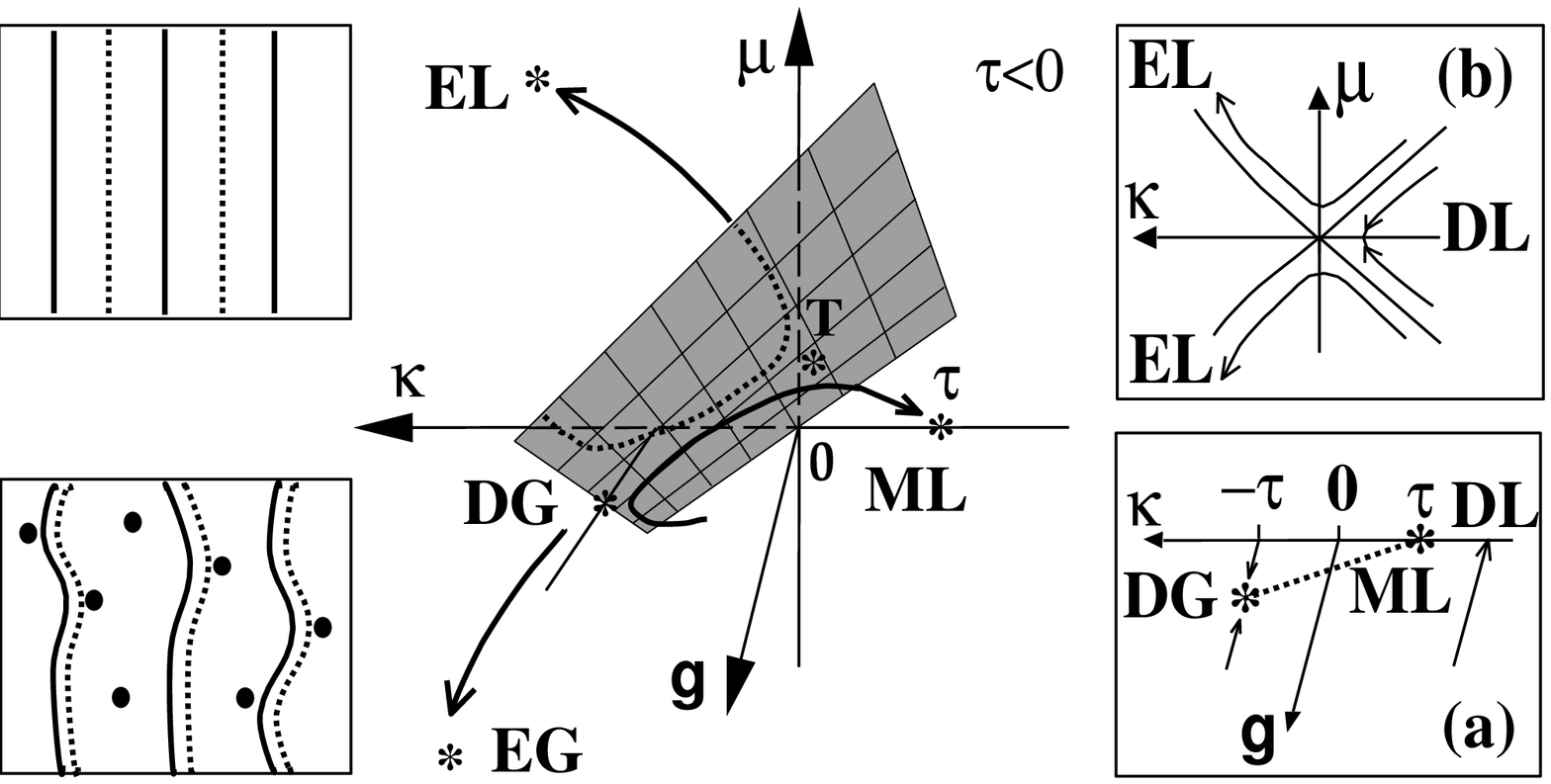}

\vspace{\baselineskip} 
 { FIG 2: RG-flow for identical disorder $(g_\mu = g, \Delta= \Delta_\mu)$. 
Inlets (a) and (b) describe the flow for the sub-case with no inter-species 
coupling ($\mu = 0$) and no disorders ($g= 0$), respectively.}
\end{figure}

Our task is to obtain the flow diagram of the full system with nonzero (but
initially small) disorder $g$ and interaction $\mu $. We shall focus on two
choices of disorder which are of particular interest: (i) identical disorder
for the two species ($g_\mu =g$, ${\delta }=0$), and (ii) completely
uncorrelated disorder ($g_\mu =0$). The case of identical disorder is the
generic situation for steps on reconstructed crystal surfaces with disorder.
It can also be constructed for the two layers of Josephson vortex lines. The
most striking feature of the RG flow in the low temperature ($\tau <0$)
regime [Fig.~2] is the strong instability of the DG fixed point with respect
to inter-species interaction $\mu \neq 0$. An attractive interaction ($\mu <0
$) favors the two species to lock into the same configuration, i.e.\ $\phi
_1=\phi _2$. Once locked, the system acts effectively as a single species
with a doubled elastic constant $K$, or equivalently a lower effective
temperature $\tau _{{\rm eff}}<\tau $, so that the effective single species
problem is in the glass phase. 
Correspondingly, (\ref{RG}) yields an RG flow away
from the unstable DG fixed point to a sink with strong inter-species
coupling {\em and} strong disorder [$g^{*},{\kappa }^{*},-\mu
^{*}\rightarrow O(1)$], for all $\tau < 0$. 
We refer to this as the elastically-coupled glass (EG)
phase. Because opposite steps are tightly bound, the EG phase is flat.
However, the domain structure is disordered (as in DG)
due to the pinning of the bound pairs of steps (or domain walls).
For $\tau >0$, we find the RG-flow to be dominated by a KT-like transition 
in the parameters $(\kappa,\mu)$, similar to that of the disorder-free 
sub-problem, with the strongly-coupled sink being the low temperature EG.

A repulsive interaction ($\mu > 0$) competes with fluctuations in the
random potential, which still attempts to lock the two species into the same
configuration. At low temperatures ($\tau < 0$), 
this competition leads to two RG sinks separated by a
second-order phase transition: If the repulsive interaction dominates, the
two species avoid each other by locking into a configuration with $\phi
_1=\phi _2+\pi $, i.e., with one species displaced by half a line spacing
with respect to the other. Such a configuration can be interpreted again as
a single species, but now with a doubled line density. This leads to a
{\em higher} effective temperature $\tau _{{\rm eff}}>0$, such that the
effective single species system is {\em not} glassy, with $g^{*}=0$.
Corresponding
to this scenario, we find for weak bare disorder a RG-flow away from DG
towards the fixed point EL, and the RG trajectories approach their pendants
in the disorder-free sub-problem. If disorder dominates, however, we obtain
a RG flow from DG towards the fixed point ML, since the disorder weakens the
inter-species coupling $\mu $ and $\kappa $, while the coupling $\mu $ in
turn weakens the disorder $g$. The phase transition separating the EL-phase
and the ML-phase is governed by an unstable fixed point (T) at 
$({\kappa }^{*},g^{*},\mu ^{*})=(-1,2,4)\left| \tau \right| /7$, 
which is the attractor of
the plane of separatrix $g\simeq \mu +2{\kappa }$.
The phase transition is second order for $\tau < 0$, with an
algebraically diverging correlation length (characterized by an exponent $%
\nu =2\left| \tau \right| /7$) upon crossing the separatrix. It is a
remarkable feature of this system that while both RG sinks are (at least
marginally) {\em disorder-free}, the unstable fixed point governing the
transition is {\em disorder-dominated}, as manifested by
$g^* > 0$ and a disordered domain structure.

The above analysis of the RG flow can be straightforwardly turned into a
phase diagram. We define an inter-species interaction energy $%
U=\mu T$, and present the phase diagram in the $(U,T)$ space for a fixed
disorder ($g$) [see Figs.~3]. The super-rough DG phases exist at $U=0$ below
a critical temperature $T_c$ given by $K_c=1/(4\pi )$. These phases are marked
 by the thick wavy line in Fig.~3(a). 
With attractive interaction $\left( U<0\right)$, there exists 
at low temperature a {\em flat, disordered} phase (EG) 
where opposite steps are bound and pinned by disorders, 
with a KT-like transition (dashed line in Fig.~3(a)) 
to a {\em logarithmically-rough, pure} phase (DL) where steps unbind.
A rather different scenario 
is obtained for repulsive interactions $(U>0)$:
The separatrix $g=\mu+2\kappa$ separates the flat, pure phase (EL) at low 
temperature and large repulsion from the two logarithmic phases. At very high
temperatures ($T \gg T_c$), the system is in the pure decoupled phase (DL), 
with the usual logarithmic interface roughness and quasi-long-range domain
order. Upon lowering the temperature beyond
the line $\tau = 0$ (thin solid line), the system settles into the ML phase
for weak repulsive interaction (compared to the strength of disorders).
The ML phase is in practice quite similar
to the high temperature phase, except for differences in the coefficient of
the logarithmic interface roughness and the exponent of the algebraic decay
of domain order. 
Further lowering the temperature beyond the separatrix (the thick solid line),
the system makes a second-order transition from the ML to the EL phase. Note
that because the critical properties there are controlled by the fixed point 
T which depends on $\tau$, the critical exponents governing this
transition actually vary {\em continuously} along the thick solid line. 
The second-order transition terminates at a point where the
separatrix intersects the line $\tau=0$ (the open circle in Fig.~3(a)). 
The transition between DL and EL at higher temperatures still occur along
the separatrix (dashed line) and is of the KT-type.

\begin{figure}
\epsfxsize=3.2truein
\hskip 0.0truein \epsffile{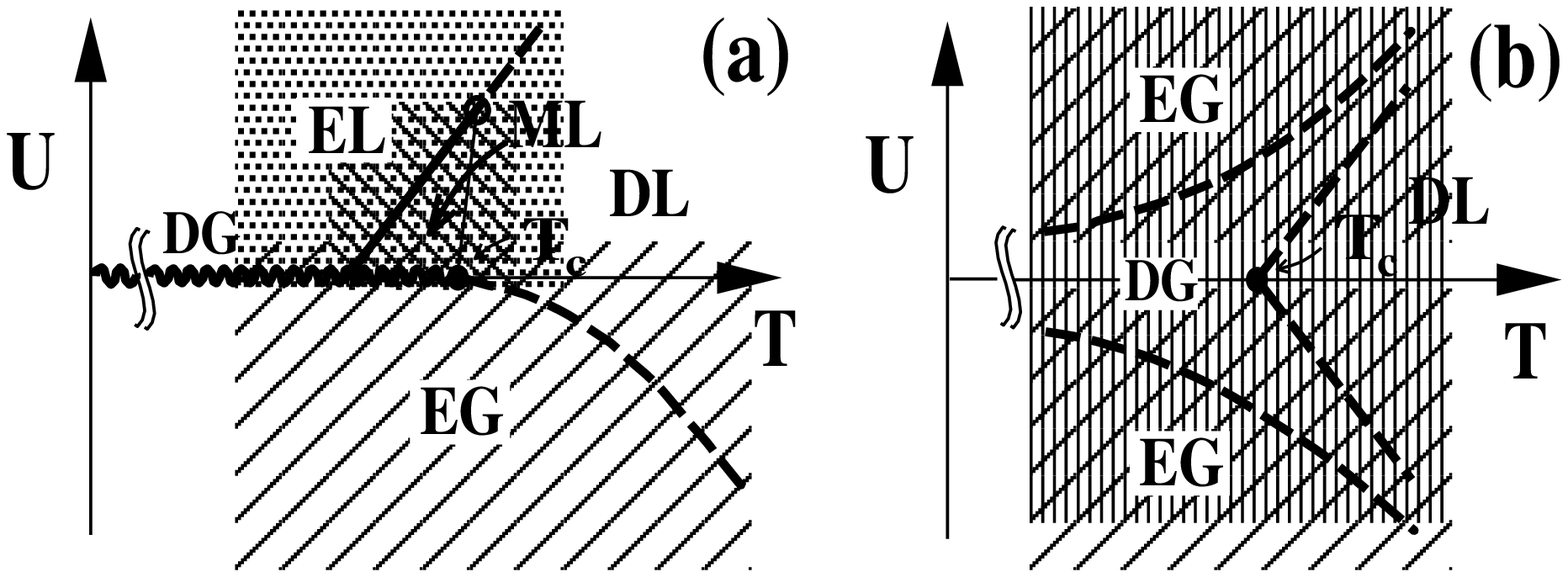}

\vspace{\baselineskip} 
{ FIG 3: Phase diagram for two species of lines in (a) identical  
and (b) uncorrelated random media.}
\end{figure}

A curious result of the above analysis is the generic {\em disappearance} 
of the glass order (as reflected by super-roughening and/or disordered
domain structures)  upon the introduction of repulsive interactions.
Such glass order was found previously for a single line array in 
random media using the replica-symmetric RG method~\cite{sr,co,gh}, 
but was absent in more recent studies using a variational method with 
replica-symmetry breaking (RSB)~\cite{k,gld}. The latter, which 
finds instead the ML phase, appears naively
consistent with our findings. On the other hand, our result can 
be used to question the internal consistency of the RSB scheme: As described 
in~Ref.~\cite{rsb}, a physical way of probing the existence of RSB 
is to take two {\em physical} replicas of a system (in identical random 
potential), and monitor the response 
to a small repulsion between the replicas. If there is a degeneracy of
low free energy states (which the RSB scheme attempts to describe),
then an infinitesimal repulsion between the replicas will force the two
to occupy different states which have {\em similar glassy properties} 
and little overlap. 
The system we have analyzed so far
can be interpreted as two physical replicas in the same random
potential. In our case, we note that the
repulsion has so strong an effect that it yields {\em  completely different
glassy properties}, i.e., from DG to ML. This indicates that the
glass order for the single line array is extremely fragile~\cite{gld2,jkf}, 
making it quite 
different from the usual scenario expected of stable (zero-temperature) glass
phases. Thus from the view point of the replica-symmetric RG analysis, 
the absence of glass order from the    
solution of the variational treatment is not surprising,
as it may be the result of subtle interactions
introduced by the RSB scheme itself.

We continue with a short discussion of the case where the disorder potentials
acting on the two species are uncorrelated, i.e. with the bare $g_\mu =0$.
This is the generic case for two layers of Josephson vortex lines in
disordered planar Josephson junctions.
Uncorrelated disorder tends to {\em decouple} the two species of lines
and competes with the locking effect of the inter-species interaction.
(Here attractive and repulsive interactions are qualitatively similar,
up to a relative $\pi$-phase shift between the two species.)
The RG analysis is more complicated than before, because one has to consider 
the full set of equations in (\ref{RG}). 
[Note that $g_\mu $ is generated by $\mu $ and $g$.] 
The basic features of the phase diagram can
be obtained by observing that the RG-flow is dominated by two KT
transitions which can be found in two sub-problems of (\ref{RG}): (i) The KT
transition of the disorder-free sub-problem. (ii)  The KT transition 
which occurs in the space  $(g_\mu,{\delta })$, when $g=\mu =\kappa+\tau=0$.
The latter has a critical separatrix $g_\mu/\sqrt{2} ={\delta }+2\tau$ which
isolates two regions of flow to a sink with $g_\mu =0$ and ${\delta }>0$, 
and another sink where $g_\mu $ grows. 
The first sink is consistent with a decoupled glass (DG)
phase as the eigenvalue for the flow of $\mu $ becomes negative there, 
while the second sink is consistent 
with the elastically-coupled glass (EG) phase where $|g_\mu|\rightarrow g$. 
The form of this separatrix for the regime of  physical initial conditions
(e.g., $g, \mu \gg g_\mu$)  is not known analytically. However, 
we determined it numerically to be well described by the form
\begin{equation}
\mu \simeq cg,  \label{crit}
\end{equation}
with the numerical constant $c\approx 0.5$. This condition, combined with
with sub-problem (i) leads to the phase diagram depicted in Fig.~3(b).
It is interesting to note that the condition (\ref{crit}) 
separating the decoupled and coupled glass phases is 
very similar, including the numerical value of $c$, to the result obtained recently for a problem involving {\em many}
layers of vortex lines~\cite{manyl}.

In conclusion, we have presented a detailed RG analysis for a model of two
interacting planar line arrays in random media. Among the findings are a
novel second-order phase transition with continuously varying critical
exponents, a disordered flat phase, and the replacement of the super-rough
glass phase by a marginally-coupled phase. 
These findings along with the structure
of the proposed phase diagrams should be accessible by experimental or
numerical investigations.

We are grateful to helpful discussions with T. Nattermann and T. Giamarchi.
This work is supported by ONR-N00014-95-1-1002 through the ONR Young
Investigator Program. TH acknowledges additional support by the A.~P.~Sloan
Foundation.

{\em Note added}: After the completion of this work, we received a preprint by
Carpentier {\it et al}~\cite{cdg} 
who obtained the same RG equations for the two
species of interacting lines as our Eq.~\ref{RG}. The analysis for the case
of uncorrelated disorders $(g_\mu =0)$ described in Ref.~\cite{cdg} agrees
qualitatively with our finding. The case of identical disorders ($g_\mu = g$
and $\delta = 0$) was however not discussed in Ref.~\cite{cdg}.

\end{multicols}


\begin{references}
\bibitem{int}  An discussion of the effect of weakly long-ranged
interactions can be found in Ref.~\cite{bk}.
\end{references}
\end{document}